\documentclass[12pt]{article}

\usepackage{amsmath,amssymb,graphics,psfrag,amsthm,cite,algorithmic}
\usepackage{amsmath,amsbsy}

\newcommand{\etc}{{\em etc}}
\newcommand{\eg}{{\em e.g.}}
\newcommand{\iid}{i.i.d.}

\newcommand{\secref}[1]{Section~\ref{#1}}
\newcommand{\figref}[1]{Fig.~\ref{#1}}
\newcommand{\tabref}[1]{Tab.~\ref{#1}}

\newtheorem{thrm}{Theorem}
\newtheorem{prop}{Proposition}

\newtheorem{corol}{Corollary}

\DeclareMathAlphabet{\mathbsf}{OT1}{cmss}{bx}{n}% bold sans serif
\DeclareMathAlphabet{\mathssf}{OT1}{cmss}{m}{sl}% slanted sans serif
\DeclareMathAlphabet{\mathcsf}{OT1}{cmss}{sbc}{n}% condensed sans serif

\DeclareSymbolFont{bsfletters}{OT1}{cmss}{bx}{n}  
\DeclareSymbolFont{ssfletters}{OT1}{cmss}{m}{n}
\DeclareMathSymbol{\bsfGamma}{0}{bsfletters}{'000}
\DeclareMathSymbol{\ssfGamma}{0}{ssfletters}{'000}
\DeclareMathSymbol{\bsfDelta}{0}{bsfletters}{'001}
\DeclareMathSymbol{\ssfDelta}{0}{ssfletters}{'001}
\DeclareMathSymbol{\bsfTheta}{0}{bsfletters}{'002}
\DeclareMathSymbol{\ssfTheta}{0}{ssfletters}{'002}
\DeclareMathSymbol{\bsfLambda}{0}{bsfletters}{'003}
\DeclareMathSymbol{\ssfLambda}{0}{ssfletters}{'003}
\DeclareMathSymbol{\bsfXi}{0}{bsfletters}{'004}
\DeclareMathSymbol{\ssfXi}{0}{ssfletters}{'004}
\DeclareMathSymbol{\bsfPi}{0}{bsfletters}{'005}
\DeclareMathSymbol{\ssfPi}{0}{ssfletters}{'005}
\DeclareMathSymbol{\bsfSigma}{0}{bsfletters}{'006}
\DeclareMathSymbol{\ssfSigma}{0}{ssfletters}{'006}
\DeclareMathSymbol{\bsfUpsilon}{0}{bsfletters}{'007}
\DeclareMathSymbol{\ssfUpsilon}{0}{ssfletters}{'007}
\DeclareMathSymbol{\bsfPhi}{0}{bsfletters}{'010}
\DeclareMathSymbol{\ssfPhi}{0}{ssfletters}{'010}
\DeclareMathSymbol{\bsfPsi}{0}{bsfletters}{'011}
\DeclareMathSymbol{\ssfPsi}{0}{ssfletters}{'011}
\DeclareMathSymbol{\bsfOmega}{0}{bsfletters}{'012}
\DeclareMathSymbol{\ssfOmega}{0}{ssfletters}{'012}

\newcommand{\genericRV}[1]{\mathssf{#1}} % generic random variable
\newcommand{\genericT}[2]{#1\left[#2\right]} % generic element #1
\newcommand{\genericTSub}[3]{\genericT{#1}{\ItoJ{#2}{#3}}} % generic
\newcommand{\src}{x}

\newcommand{\rvSrc}{\genericRV{\src}}

\newcommand{\rvSrcT}[1]{\genericT{\rvSrc}{#1}}

\newcommand{\rvSrcTSub}[2]{\genericTSub{\rvSrc}{#1}{#2}}

\newcommand{\srcT}[1]{\genericT{\src}{#1}}
\newcommand{\quant}{\hat{\src}}

\newcommand{\rvQuant}{\genericRV{\quant}}

\newcommand{\rvQuantT}[1]{\genericT{\rvQuant}{#1}}
\newcommand{\rvQuantTSub}[2]{\genericTSub{\rvQuant}{#1}{#2}}

\newcommand{\sinfoDec}{w}
\newcommand{\rvSinfoDec}{\genericRV{\sinfoDec}}

\newcommand{\rvSinfoDecT}[1]{\genericT{\genericRV}{\sinfoDec}}

\newcommand{\mesg}{m}
\newcommand{\mesgT}[1]{\genericT{\mesg}{#1}}
\newcommand{\mesghT}[1]{\genericT{\hat{\mesg}}{#1}}

\newcommand{\srcAlph}{{\mathcal{\MakeUppercase{\src}}}}

\newcommand{\sinfoDecAlph}{{\mathcal{\MakeUppercase{\sinfoDec}}}}

\usepackage{euscript}

\newcommand{\pcond}{p_{\rvQuant|\rvSrc}}
\newcommand{\pjoint}{p_{\rvQuant,\rvSrc}}
\newcommand{\pquant}{p_{\rvQuant}}
\newcommand{\psrc}{p_{\rvSrc}}
\newcommand{\frdfo}[1]{R_{\mathrm{f}}(#1)}
\newcommand{\frdfi}[1]{R^{(I)}_{\mathrm{f}}(#1)}
\newcommand{\genericLength}{m}
\newcommand{\FFDelay}{\Delta}

\newcommand{\compSimp}[1]{\EuScript{C}\left(#1\right)}
\newcommand{\decompSimp}[1]{\EuScript{C}^{-1}\left(#1\right)}
\newcommand{\lengthFunc}[1]{\EuScript{L}\left(#1\right)}

\newcommand{\compress}[1]{\EuScript{C}_{\rvQuant}\left(#1\right)}
\newcommand{\decompress}[1]{\EuScript{C}^{-1}_{\rvQuant}\left(#1\right)}
\newcommand{\shaper}[1]{\EuScript{S}_{\rvQuant|\rvSrc}\left(#1\right)}
\newcommand{\deshaper}[1]{\EuScript{S}^{-1}_{\rvQuant|\rvSrc}\left(#1\right)}
\newcommand{\blockSize}{n}
\newcommand{\nominalBlockSize}{\mathtt{N}}
\newcommand{\minBlockSize}{\mathtt{M}}

\newcommand{\dmax}{d_{\max}}
\newcommand{\numStages}{K}
\newcommand{\hFactor}{\frac{H(\rvQuant)}{H(\rvQuant|\rvSrc)}}
\newcommand{\bhFactor}{\left[\hFactor\right]}
\newcommand{\hFactorInverse}{\frac{H(\rvQuant|\rvSrc)}{H(\rvQuant)}}

\newcommand{\ffencoder}[1]{f(#1)}
\newcommand{\ffdecoder}[1]{f^{-1}(#1)}

\renewcommand{\genericT}[2]{#1_{#2}}
\renewcommand{\genericTSub}[3]{#1_{#2}^{#3}}

\newcommand{\defeq}{\stackrel{\Delta}{=}}

\newcommand{\dist}[2]{d(#1,#2)}

\newcommand{\binEnt}[1]{H_b(#1)}
\newcommand{\rate}{R}

\begin{document}

\newlength{\cheatLength}
\setlength{\cheatLength}{-.3in}
\renewcommand{\textfraction}{0}

\title{Source Coding with Fixed Lag Side Information}
\author{\normalsize
Emin Martinian and Gregory W. Wornell\\
[-5pt] \small Massachusetts Institute of Technology  \\
[-5pt] \small Cambridge, MA 02139\\
[-5pt] \small \{emin,gww\}@allegro.mit.edu}

\date{}

\maketitle

\thispagestyle{myheadings}
\markboth{Proceedings of the 42nd Annual Allerton Conference, (Monticello, IL)
  2004}{Proceedings of the 42nd Annual Allerton Conference, (Monticello, IL)
  2004}
%\pagenumbering{gobble}
\pagestyle{plain}

\vspace{-.25in}
\begin{abstract}
We consider source coding with fixed lag side information at the
decoder.  We focus on the special case of perfect side information
with unit lag corresponding to source coding with feedforward (the
dual of channel coding with feedback) introduced by Pradhan
\cite{Pradhan:2004:isit}.  We use this duality to develop a linear
complexity algorithm which achieves the rate-distortion bound for any
memoryless finite alphabet source and distortion measure.
\end{abstract}

\section{Introduction}

There is a growing consensus that understanding complex, distributed
systems requires a combination of ideas from communication and control
\cite{Murray:2003:icsm}.  Adding communication 
constraints to traditional control problems or adding real-time
constraints to communication problems
has recently yielded interesting results
\cite{borkar:2001:siamjco,Elia:2001:cnt,sahai:2000:phd,tatikonda:2000:phd,Wong:1997:cnt}.
We consider a related aspect of this interaction by exploring the possible
advantages that the feedback/feedforward in control
scenarios can provide in compression.  Specifically, we explore a
variant of the Wyner-Ziv problem \cite{it:wyner_1975} where causal
side information about the source is available with a fixed lag to the
decoder and explore how such side information may be used.  

\begin{figure}[b]
\begin{center}
\psfrag{Source}{\Large Source $\rvSrcT{i}$}
\psfrag{Encoder}{\Large Sensor}
\psfrag{Decoder}{\Large Controller}
\psfrag{Delay}{\Large Delay}
\psfrag{Compressed}{\Large Compressed data $\mesg$}
\psfrag{xd}{\Large $\rvSrcT{i-\Delta}$}
\psfrag{xh}{\Large $\rvQuantT{i}$}
\resizebox{\textwidth}{!}{\includegraphics{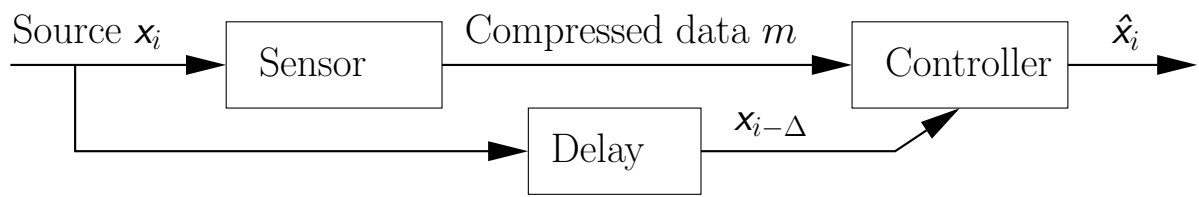}}
\caption{A sensor compresses and sends the source sequence
  $\rvSrcT{1}$, $\rvSrcT{2}$, $\ldots$, to a
  controller which reconstructs the quantized sequence $\rvQuantT{1}$,
  $\rvQuantT{2}$, $\ldots$, in order to take some control action.
  After a delay or lag of $\Delta$, the controller observes the 
  original, uncompressed data directly.
\label{fig:cs}}
\end{center}
\end{figure}

For example, consider a remote sensor that sends its observations to a
controller as illustrated in \figref{fig:cs}.  The sensor may be a
satellite or aircraft reporting the upcoming temperature, wind speed,
or other weather data to a vehicle.  The sensor observations must be
encoded via lossy compression to conserve power or bandwidth.  In
contrast to the standard lossy compression scenario, however, the
controller directly observes the original, uncompressed data after
some delay.  The goal of the sensor observations is to provide the
controller with information about upcoming events {\em before} they
occur.  Thus at first it might not seem that observing the true,
uncompressed data {\em after} they occur would be useful.  Our main
goal is to try to understand how these delayed observations of the
source data (which we call side information) can be used.  Our main
result is that such information can be quite valuable.

The following toy problem helps illustrate some relevant issues.
Imagine that Alice plays a game where she will be
asked 10 Yes/No questions.  Of these questions, 5 have 
major prizes while the others have minor prizes.  After answering
each question, she is told the correct answer as well as what the
prize for that question is and receives the prize if she is correct.
Bob knows all the questions and the corresponding prizes beforehand
and wishes to help Alice by preparing a ``cheat-sheet'' for her. But Bob
only has room to record 5 answers.  Is there a cheat-sheet encoding
strategy that guarantees that Alice will always correctly answer the
questions with the 5 best prizes?
No such strategy exists using a classical compression scheme.
Instead, as illustrated in \secref{sec:simple-example}, the optimal
strategy requires an encoding which uses the fact 
that Alice gains information about the prize {\em after} answering.

In the rest of the paper, we study the fixed lag side information
problem.  Since solving the general problem seems difficult, we begin
by focusing on perfect side information with a unit lag.  This special
case is the feedforward source coding problem introduced by Pradhan
\cite{Pradhan:2004:isit} and is dual (in the sense of
\cite{Pradhan:2003:it,Barron:2003:it}) to channel coding with
feedback.
By using the feedforward side information, it is possible to construct
low complexity source coding systems which can achieve the
rate-distortion bound.  Specifically, \cite{Pradhan:2004:isit}
describes how to adapt the Kailath-Schalkwijk scheme for the Gaussian
channel with feedback \cite{Schalkwijk:1966:it} to the Gaussian source
squared distortion scenario with feedforward side information.  In
this paper, we consider finite alphabet sources with arbitrary
memoryless distributions and arbitrary distortion measures.  Since Ooi
and Wornell's channel coding with feedback scheme \cite{Ooi:1998:it}
achieves the best error exponent with minimum complexity, we
investigate the source coding dual.  Specifically, we show that the
source coding dual of the Ooi--Wornell scheme achieves the
rate-distortion bound with linear complexity.

We begin by describing the problem in
\secref{sec:problem-description}.  Next we present a simple example of
how feedforward side information can be useful in the binary erasure
quantization problem in \secref{sec:simple-example}.  In
\secref{sec:complicated-example}, we consider the more complicated
example of quantizing a binary source with respect to Hamming
distortion.  We present our source coding algorithm for general
sources in \secref{sec:gener-feedf-compr} and show that it achieves
the rate-distortion function with low complexity.  We
close with some concluding remarks in \secref{sec:concluding-remarks}.

\section{Problem Description}
\label{sec:problem-description}

Random variables are denoted using the sans serif
font (\eg, $\genericRV{x}$) with deterministic values using serif
fonts (\eg, $x$).  We represent the $i$th element of a sequence as
$\genericT{x}{i}$ and denote a subsequence including elements from $i$
to $j$ as $\genericTSub{x}{i}{j}$.

We consider (memoryless) source coding with fixed lag side
information and represent an instance of the problem with the tuple
$(\srcAlph,\sinfoDecAlph,p_{\rvSrc,\rvSinfoDec},\dist{\cdot}{\cdot},\FFDelay)$
where $\srcAlph$ and $\sinfoDecAlph$ represent the source and side
information alphabets, $p_{\rvSrc,\rvSinfoDec}(\src,\sinfoDec)$ represents
the source and side information joint distribution,
$\dist{\cdot}{\cdot}$ represents the distortion measure, and $\FFDelay$
represents the delay or lag.  Specifically, the source and side
information each consist of a sequence 
of $\blockSize$ random variables $\rvSrcTSub{1}{\blockSize}$ and
$\genericTSub{\rvSinfoDec}{1}{\blockSize}$ taking values in
$\srcAlph$ and $\sinfoDecAlph$ generated according to the
distribution
$p_{\rvSrcTSub{1}{\blockSize},\genericTSub{\rvSinfoDec}{1}{\blockSize}}(\genericTSub{\src}{1}{\blockSize})
= \prod_{i=1}^n p_{\rvSrc,\rvSinfoDec}(\srcT{i},\genericT{\sinfoDec}{i})$.

A rate $\rate$ encoder, $\ffencoder{\cdot}$, maps
$\rvSrcTSub{1}{\blockSize}$ to a bit sequence represented as an
integer $\mesg \in \{1, 2, \ldots, 2^{n \rate}\}$.  The corresponding
decoder $\ffdecoder{\cdot}$ works as follows.  At time $i$, the decoder
takes as input $\mesg$ as well as the side information samples,
$\genericTSub{\rvSinfoDec}{1}{i-\FFDelay}$, and produces the $i$th reconstruction
$\rvQuantT{i}$.  A distortion of $\dist{\rvSrcT{i}}{\rvQuantT{i}}$ is
then incurred for the $i$th sample where $\dist{\cdot}{\cdot}$ is a
mapping from $\srcAlph\times\srcAlph$ to the interval $[0,\dmax]$.

The basic problem can be specialized to the original (non-causal)
Wyner-Ziv problem \cite{it:wyner_1975}, by allowing a negative delay
$\FFDelay=-\infty$.  Similarly, setting $\FFDelay=0$ yields a causal
version of the Wyner-Ziv problem.  Finally, letting the side
information be exactly the same as the source with a positive delay
yields the feedforward source coding problem studied in
\cite{Pradhan:2004:isit}.  For all these cases, the goal is to
understand the fundamental rate-distortion-complexity performance.  To
show that the benefits of fixed lag side information are worth
investigating, we focus on the feedforward case where
$\rvSinfoDec=\rvSrc$ with unit delay $\FFDelay=1$ throughout the rest of this
paper.

For memoryless sources, the information feedforward rate-distortion
function, $\frdfi{D}$, is defined to be the same as Shannon's classical
rate-distortion function: 
\begin{equation}
\label{eq:frdfi-def}
\frdfi{D} = \inf_{p_{\rvQuant|\rvSrc}: E[\dist{\rvSrc}{\rvQuant}]\leq
  D} I(\rvQuant;\rvSrc).
\end{equation}
The operational feedforward rate-distortion function, $\frdfo{D}$, is
the minimum rate required such that there exists a sequence
of encoders and decoders with average distortion,
$\frac{1}{\blockSize}\sum_{i=1}^{\blockSize}d(\rvSrcT{i},\rvQuantT{i})$,
asymptotically approaching $D$.  As observed in
\cite{Pradhan:2004:isit} and shown in the appendix, the information
and operational feedforward rate-distortion functions are the same.
Thus feedforward does not reduce the rate required, but as we argue in
the rest of this paper, it allows us to approach the rate-distortion
function with low complexity.

\section{Example: Binary Source \& Erasure Distortion}
\label{sec:simple-example}

\begin{figure}[htb]
\begin{center}
\psfrag{increaseii}{\LARGE increase i}
\psfrag{Start}{\LARGE Start}
\psfrag{setieq1}{\LARGE $i=1$}
\psfrag{Yes}{\LARGE Yes}
\psfrag{No}{\LARGE No}
\psfrag{Send}{\LARGE Send $\mesgT{i}$}
\psfrag{mito}{\LARGE through}
\psfrag{tchannel}{\LARGE channel}
\psfrag{Getnext}{\LARGE $z=$ next}
\psfrag{fchannel}{\LARGE channel}
\psfrag{sample}{\LARGE sample}
\psfrag{yierased}{\LARGE $z$ erased?}
\psfrag{mhieqx}{\LARGE $\mesghT{i} = z$}
\psfrag{input}{\LARGE channel}
\psfrag{erased}{\LARGE erased?}
\resizebox{4.5in}{!}{\includegraphics{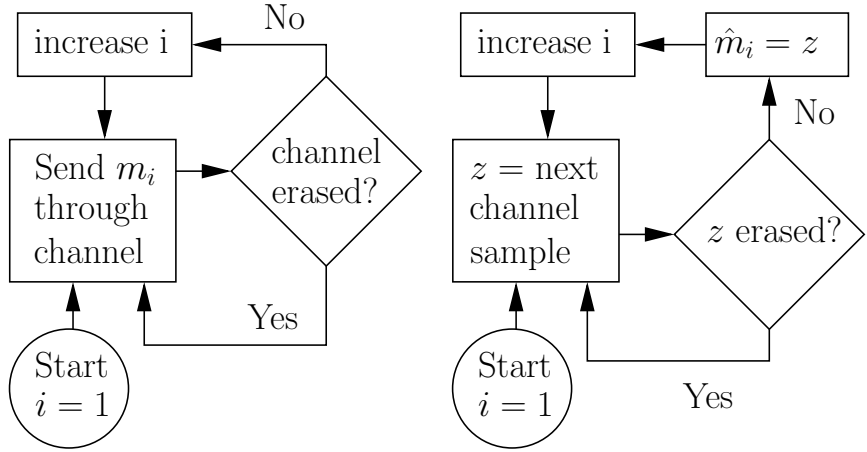}}
\caption{Encoder (left) for transmitting a message $m=m_1,m_2,\ldots$
  across an erasure channel with feedback and decoder (right) for
  producing an estimate of the transmitted message $\hat{m}$.
\label{fig:becf}}
\end{center}
\vspace{\cheatLength}
\end{figure}

The simplest example in channel coding with feedback is the erasure
channel and in this case the algorithm in \figref{fig:becf}
achieves capacity.  At time 1 the encoder puts message bit $\mesgT{1}$
into the  
channel.  If it is received correctly, the transmitter then transmits
$\mesgT{2}$, otherwise $\mesgT{1}$ is repeated until it is
successfully received.  The 
same process is used for $\mesgT{2}$, $\mesgT{3}$, etc.  For example,
to send the 
message $0 1 0 1$ though a channel where samples 2, 3, 6, and 7 are
erased, the transmitter would send $0 1 1 1 0 1 1 1$ and the receiver
would see  $0 * * 1 0 * * 1$.  In general, if there are $\blockSize$ message
bits, $\mesgT{1}$, $\mesgT{2}$, $\ldots$, $\mesgT{\blockSize}$, and
$e$ erasures, then exactly 
$\blockSize+e$ channel uses are required.  This yields a transmission rate of
$\blockSize/(\blockSize+e)$ which is exactly the channel capacity.

\begin{figure}[htb]
\begin{center}
\psfrag{increaseii}{\LARGE increase i}
\psfrag{Start}{\LARGE Start}
\psfrag{setieq1}{\LARGE $i=1$}
\psfrag{Yes}{\LARGE Yes}
\psfrag{No}{\LARGE No}
\psfrag{Send}{\LARGE Set next}
\psfrag{mito}{\LARGE sample}
\psfrag{tchannel}{\LARGE to $\mesgT{i}$}
\psfrag{Getnext}{\LARGE $z=$ next}
\psfrag{fchannel}{\LARGE source}
\psfrag{sample}{\LARGE sample}
\psfrag{yierased}{\LARGE $z$ erased?}
\psfrag{mhieqx}{\LARGE $\mesgT{i} = z$}
\psfrag{true}{\LARGE true}
\psfrag{source}{\LARGE source}
\psfrag{erased}{\LARGE erased?}
\resizebox{4.5in}{!}{\includegraphics{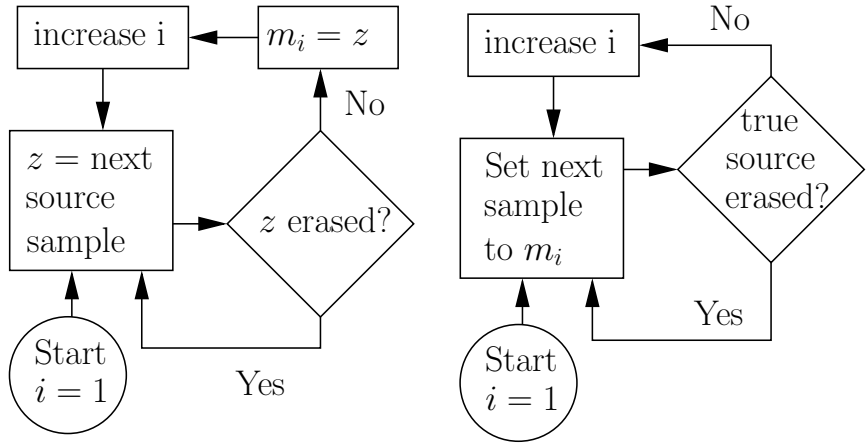}}
\caption{Encoder (left) and decoder (right) for the binary erasure
  quantization problem which is dual to the binary erasure channel.
\label{fig:beqf}}
\end{center}
\vspace{\cheatLength}
\end{figure}

The dual to the binary erasure channel (BEC) is the binary erasure
quantization problem (BEQ).  In the BEQ, each source sample can be
either 0, 1, or * where * represents ``don't care''.  The distortion
measure is such that 0 and 1 cannot be changed but * can be quantized
to either 0 or 1 with no distortion.  The BEQ models the game
introduced in the introduction.\footnote{Yes/No answers for questions
  with major prizes map to 1/0 values for the source while questions
  with minor prizes map to a value of * for the source.  The
  distortion measure represents the restriction that questions with
  major prizes must be answered correctly while the answers for the
  other questions are irrelevant.}
To develop a source coding with feedforward algorithm for the
BEQ, we can dualize the channel coding with feedback algorithm for the
BEC as illustrated in \figref{fig:beqf}.

Assume the source is $\rvSrcTSub{1}{8} = 0 * * 1 0 * * 1$.  The
encoder compresses this to $\mesg = 0 1 0 1$ by ignoring all the *
symbols and sends this to the receiver.  At time 1, the receiver
chooses $\rvQuantT{1}$ to be the first bit in the encoding (i.e.,
$\rvQuantT{1} = \mesgT{1} = 0$).  From the feed-forward the receiver
realizes this is correct after it makes its choice.  Next, the
receiver chooses $\rvQuantT{2}$ to be the next bit received (i.e.,
$\rvQuantT{2} = \mesgT{2} = 1$).  After choosing this reconstruction,
the receiver is told that in fact $\rvSrcT{2} = *$ so even though
$\rvSrcT{2} \neq \rvQuantT{2}$, no distortion is incurred.  At this
point, the receiver realizes that $\mesgT{2}$ must have been intended
to describe something after $\rvSrcT{2}$.  So at time 3, the receiver
chooses $\rvQuantT{3} = \mesgT{2} = 1$.  Once again the receiver
learns that this is incorrect since $\rvSrcT{3} = *$, but again no
penalty is incurred.  Again, the receiver decides that $\mesgT{2}$
must have been intended to describe something else so it chooses
$\rvQuantT{4} = \mesgT{2} = 1$ at time 4.  This turns out to be
correct and so at time 5 the receiver chooses $\rvQuantT{5} =
\mesgT{3}$, etc.

In the encoder/decoder described above, the encoder sends the
non-erased bits of $\rvSrcTSub{1}{\blockSize}$ and the decoder tries
to match up the compressed data to the source.  This system yields distortion 0
provided that at least $\blockSize-e$ bits are sent where $e$ denotes
the number of * symbols in the source vector.  It is straightforward
to show that no encoder/decoder can do better for any value of
$\blockSize$ or $e$.  A system not taking advantage of the feedforward could
asymptotically achieve the same performance but it would require more
complexity and more redundancy.  Thus just as in the erasure channel
with feedback, we see that for the erasure source, feedforward allows
us to achieve the minimum possible redundancy with minimum complexity.

\section{Example: Binary Source \& Hamming Distortion}
\label{sec:complicated-example}

The example in \secref{sec:simple-example} illustrates that
feedforward can be useful in source coding by using some special
properties of the BEQ problem.  Next, we consider a somewhat more
complicated example to illustrate that a key idea in developing {\em
lossy} compression algorithms for source coding with feedforward is
the use of classical {\em lossless} compression algorithms.
Specifically, we consider a binary source which is equally likely to
be either zero or one, and we consider the Hamming distortion measure
$d(\src,\quant) = |\src-\quant|$.  As is well known, the
rate-distortion function for this case is $R(D)=1-\binEnt{D}$ where
$\binEnt{\cdot}$ is the binary entropy function.  In the following we
outline a scheme which achieves a distortion of $D_0=0.11$ and rate
$R(D_0) = 1 - \binEnt{0.11} \approx 1/2$.

Let $\compSimp{\cdot}$ and $\decompSimp{\cdot}$ be a lossless
compression and decompression algorithm for a Bernoulli($D_0$) source.
Specifically, $\compSimp{\cdot}$ takes as input $t$ bits with a
fraction $D_0$ ones and maps them into $\binEnt{D_0}t \approx t/2$
uniformly distributed bits while $\decompSimp{\cdot}$ maps $t'$
approximately uniformly distributed bits into $t'/\binEnt{D_0}
\approx 2 t'$ bits with a fraction $D_0$ ones.  To simplify the exposition,
we assume that for $t' \geq \minBlockSize$, these approximations are
exact.  A more careful treatment appears in
\secref{sec:gener-feedf-compr}.

The feedforward lossy compression system encoder takes a sequence of
source samples, $\rvSrcTSub{1}{\blockSize}$, where $\blockSize =
\minBlockSize (2^{\numStages}-1)$ and encodes by producing the
following codewords:
\begin{align}
b_1 & \defeq \rvSrcTSub{\blockSize-\minBlockSize+1}{\blockSize}\\
b_2 &
\defeq
\rvSrcTSub{\blockSize-3\minBlockSize+1}{\blockSize-\minBlockSize} \oplus
\decompSimp{b_1} =
\rvSrcTSub{\blockSize-3\minBlockSize+1}{\blockSize-\minBlockSize}
\oplus
\decompSimp{\rvSrcTSub{\blockSize-\minBlockSize+1}{\blockSize}}\\
b_3
& \defeq 
\rvSrcTSub{\blockSize-7\minBlockSize+1}{\blockSize-3\minBlockSize}
 \oplus \decompSimp{b_2}
 =
\rvSrcTSub{\blockSize-7\minBlockSize+1}{\blockSize-3\minBlockSize}
 \oplus \decompSimp{\rvSrcTSub{\blockSize-3\minBlockSize+1}{\blockSize-\minBlockSize} \oplus
\decompSimp{b_1}} \\
&= \rvSrcTSub{\blockSize-7\minBlockSize+1}{\blockSize-3\minBlockSize} 
 \oplus \decompSimp{\rvSrcTSub{\blockSize-3\minBlockSize+1}{\blockSize-\minBlockSize}
\oplus
\decompSimp{\rvSrcTSub{\blockSize-\minBlockSize+1}{\blockSize}}}\\
\vdots & \hspace{1in} \vdots\\
\label{eq:bk}
b_{\numStages} & \defeq 
\rvSrcTSub{1}{\minBlockSize 2^{\numStages-1}}
 \oplus \decompSimp{b_{\numStages-1}}
 =
\rvSrcTSub{1}{\minBlockSize 2^{\numStages-1}}
 \oplus
 \decompSimp{\rvSrcTSub{2^{\numStages-1}\cdot\minBlockSize+1}{3\minBlockSize 2^{\numStages-1}} \oplus
\decompSimp{b_{\numStages-2}}} \\
&= \rvSrcTSub{1}{\minBlockSize 2^{\numStages-1}}
 \oplus
 \decompSimp{\rvSrcTSub{2^{\numStages-1}\cdot\minBlockSize+1}{3\minBlockSize 2^{\numStages-1}} \oplus
\decompSimp{\rvSrcTSub{3\minBlockSize
 2^{\numStages-1}+1}{7\minBlockSize 2^{\numStages-2}} \oplus \ldots
 \oplus \decompSimp{\rvSrcTSub{\blockSize-\minBlockSize+1}{\blockSize}}}}
\end{align}
according to the general rule
\begin{equation}
b_i \defeq \rvSrcTSub{\blockSize-(2^i-1)\minBlockSize +
  1}{\blockSize- (2^{i-1}-1)\minBlockSize} \oplus \decompSimp{b_{i-1}}.
\end{equation}
The output of the encoder is the $\minBlockSize \cdot
2^{\numStages-1}$ bit sequence $b_{\numStages}$ for the last block.

As we see from \eqref{eq:bk}, $b_{\numStages}$ is a description of the
first block of source samples corrupted by the addition of
$\decompSimp{b_{\numStages-1}}$.  The decoder reconstructs the first
$\minBlockSize\cdot 2^{\numStages-1}$ source samples via
\begin{equation}
\rvQuantTSub{1}{\minBlockSize 2^{\numStages-1}} \defeq b_{\numStages}
= \rvSrcTSub{1}{\minBlockSize 2^{\numStages-1}} \oplus
\decompSimp{b_{\numStages-1}}. 
\end{equation}
The distortion for this block is approximately $D_0$ since, by
assumption, the decompresser $\decompSimp{\cdot}$ maps its input to a
sequence with a fraction $D_0$ ones.  The error between the reconstruction and
the true source obtained via feedforward is a description of future
source samples shaped by the function $\compSimp{\cdot}$.  Thus, to
reconstruct the next block, the decoder uses the feedforward,
$\rvSrcTSub{1}{\minBlockSize 2^{\numStages-1}}$, to produce
\begin{equation}
\rvQuantTSub{\minBlockSize 2^{\numStages-1}+1}{3\minBlockSize
  2^{\numStages-2}} \defeq \compSimp{\rvSrcTSub{1}{\minBlockSize
  2^{\numStages-1}} \oplus b_{\numStages}} = b_{\numStages-1} =
  \rvSrcTSub{\minBlockSize 2^{\numStages-1}+1}{3\minBlockSize 
  2^{\numStages-2}} \oplus \decompSimp{b_{\numStages-2}}.
\end{equation}
Once again the distortion is approximately $D_0$ since the
decompresser maps its input to a sequence with $D_0$ ones.  

The decoder proceeds in this manner and obtains a distortion of
approximately $D_0$ for each block except the last which yields no
distortion.  The average distortion is therefore roughly $D_0$.
Since $\minBlockSize 
2^{\numStages-1}$ bits are required to describe $b_{\numStages}$ in
encoding the $\minBlockSize (2^{\numStages}-1)$ source samples, the
average bit rate is $2^{\numStages-1}/(2^{\numStages}-1)\approx 1/2$.  Thus
by taking advantage of the source feedforward, we can obtain a point
on the rate distortion curve simply by using a low complexity lossless
compression algorithm.

\section{Finite Alphabet Sources \& Arbitrary Distortion}
\label{sec:gener-feedf-compr}

In this section, we generalize the construction in
\secref{sec:complicated-example} to arbitrary rates, source
distributions and distortion measures.  We require two components: a
lossless compression/decompression algorithm and a shaping algorithm.
Using these subsystems, we describe our feedforward source coding
algorithm and present an analysis of its rate and distortion.

\subsection{Feedforward Source Coding Subsystems}

Our lossless compression and shaping algorithms must be efficient in
some sense for the overall feedforward source coding algorithm to
approach the rate-distortion function.  Instead of delving into the
details of how to build efficient compression and shaping algorithms,
we define admissible systems to illustrate the required properties.
We then describe how efficient subsystems can be combined.  

\subsubsection{Lossless Compression Subsystem}
\label{sec:lossl-compr-subsyst}

We define a $(\delta,\epsilon,\genericLength)$ admissible lossless
compression system 
as follows.  On input of
$\genericLength$ samples from which are $\delta$-strongly
typical\footnote{A sequence is $\delta$-strongly typical if the
empirical fraction of occurrences of each possible outcome differs by
at most $\delta$ from the expected fraction of outcomes and no
probability zero outcomes occur.}  according to the distribution
$\pquant$, the compressor, denoted $\compress{\cdot}$, returns
$\genericLength \cdot H(\rvQuant) + \epsilon$ bits.
If the input is not $\delta$-strongly typical, the output is
undefined.  The corresponding decompresser, $\decompress{\cdot}$ takes
the resulting bits and reproduces the original input.

\subsubsection{Shaping Subsystem}
\label{sec:shaping-subsystem}

We define a $(\delta,\epsilon,\genericLength)$ admissible shaping
system as follows. 
On input of a sequence of $\genericLength$ bits, and a semi-infinite
sequence of samples, $\rvSrcTSub{1}{\infty}$ which is
$\delta$-strongly typical according to the distribution $\psrc$, the
shaper $\shaper{\cdot}$ returns a sequence of
$\genericLength'=\genericLength\cdot[H(\rvQuant)/H(\rvQuant|\rvSrc)] +
\epsilon$ 
samples, $\rvQuantTSub{1}{\genericLength'}$, such that
$(\rvSrcTSub{1}{\genericLength'},\rvQuantTSub{1}{\genericLength'})$ is
$\delta$-strongly typical according to the distribution $\pjoint$.  If
the input is not $\delta$-strongly typical, the output is undefined.
The corresponding deshaper takes the pair of sequences
$\rvQuantTSub{1}{\genericLength'}$ and
$\rvSrcTSub{1}{\genericLength'}$ as input and returns the original
sequence of $\genericLength$ bits.

The compression and shaping systems described previously are
fixed-to-variable and variable-to-fixed systems respectively.  Hence,
for notational convenience we define the corresponding length functions
$\lengthFunc{\compress{\cdot}}$ and $\lengthFunc{\shaper{\cdot}}$ as
returning the length of their respective arguments.

\subsubsection{Efficient Shaping and Compression Systems}

We call a lossless compression system or a shaping system efficient if
both $\delta$ and $\epsilon$ can be made arbitrarily small for
$\genericLength$ large enough.  Efficient lossless compression systems
can be implemented in a variety of ways.  For example, arithmetic
coding is one well-known approach.  Perhaps less well-known is that
shaping systems can also be implemented via arithmetic 
coding \cite{Ooi:1998:it}.  Specifically, by using the decompresser
for an arithmetic code as a shaper, we can map a sequence of bits into
a sequence with an arbitrary distribution.  The compressor for the
arithmetic code takes the resulting sequence and returns the original
bit sequence.

\newcommand{\ct}{T}
\newcommand{\cl}{L}
\newcommand{\cs}{S}

\subsection{Feedforward Encoder and Decoder}

Since the encoder for our feedforward lossy compression system is
based on a variable-to-fixed shaper and a fixed-to-variable
compressor, it is a variable-to-variable system.  In practice, one
could use buffering, padding, or other techniques to account for this
when encoding a fixed length source or when required to produce a
fixed length encoding.  We do not address this issue further here.
Instead, we assume that there is a nominal source block size
parameter, $\nominalBlockSize$, and buffering, padding, look-ahead,
\etc. is used to ensure that the system encodes $\nominalBlockSize$
source samples (or possibly slightly more or less).  Also, we assume
that there is a minimum block size parameter, $\minBlockSize$, which
may be chosen to achieve an efficient shaping or lossless compression
subsystem.

Once $\nominalBlockSize$ and $\minBlockSize$ are fixed, the
feedforward encoder takes as input a stream of inputs
$\rvSrcTSub{1}{\infty}$ and encodes it as described in \tabref{tab:enc}.
The feedforward decoder takes as input the resulting bit string, $b$,
and decodes it as described in \tabref{tab:dec}.
\secref{sec:complicated-example} describes an example of the encoding
and decoding algorithm with a shaper (denoted $\decompSimp{\cdot}$)
mapping uniform bits to 
Bernoulli($0.11$) bits.  This example does not require a compressor
because the $p_{\rvQuant}$ distribution is incompressible.

\begin{table}[htb]
\caption{The Feedforward Encoder.    \label{tab:enc}} 
\vspace{-0.25in}
\begin{center}
\fbox{%
\begin{minipage}{\textwidth}
\begin{small}
\begin{algorithmic}[1]
\STATE Initialize $\ct=1$, $\cl = \minBlockSize$, and reverse the
input so that in the following
$\rvSrcTSub{1}{\blockSize}=\rvSrcTSub{\blockSize}{1}$. 
\STATE Take the block of source samples
  $\rvSrcTSub{\ct}{\ct+\cl}$ and generate a
  ``noisy version''
  $\rvQuantTSub{\ct}{\ct+\cl}$ (\eg, by
  generating each $\rvQuantT{i}$ from the corresponding $\rvSrcT{i}$
  according to $\pcond$).  
\WHILE {$\cl+\ct < \nominalBlockSize$}
\STATE Compress $\rvQuantTSub{\ct}{\ct+\cl}$ to obtain the bit sequence
  $b =
  \compress{\rvQuantTSub{\ct}{\ct+\cl}}$. 
\STATE $\ct \leftarrow \ct + \cl + 1$
\STATE $\cl \leftarrow
\lengthFunc{\shaper{b,\rvSrcTSub{\ct}{\infty}}}$.
\STATE $\rvQuantTSub{\ct}{\ct+\cl} \leftarrow
  \shaper{b,\rvSrcTSub{\ct}{\infty}}$
\ENDWHILE
\STATE {\bf return} $\compress{\rvQuantTSub{\ct}{\ct+\cl}}$
\end{algorithmic}
\end{small}
\end{minipage}
}
\end{center}
\vspace{-0.25in}
\end{table}

\vspace{-0.25in}

\begin{table}[htb]
\caption{The Feedforward Decoder. \label{tab:dec}}
\vspace{-0.25in}
\begin{center}
\fbox{%
\begin{minipage}{\textwidth}
\begin{small}
\begin{algorithmic}[1]
\STATE Initialize $\ct$ to the length of the sequence encoded in
  $b$.
\WHILE {$\ct > 1$}
\STATE $\cl \leftarrow \lengthFunc{\decompress{b}}$
\STATE $\ct \leftarrow \ct - \cl+1$
\STATE $\rvQuantTSub{\ct}{\ct+\cl} \leftarrow \decompress{b}$
\STATE Get $\rvSrcTSub{\ct}{\ct+\cl}$ via the feedforward
  information
\STATE $b \leftarrow
\deshaper{\rvQuantTSub{\ct}{\ct+\cl},\rvSrcTSub{\ct}{\ct+\cl}}$
\ENDWHILE
\STATE {\bf return} the reversed version of $\rvQuantTSub{1}{\blockSize}$
\end{algorithmic}
\end{small}
\end{minipage}
}
\end{center}
\vspace{-0.25in}
\end{table}

\subsection{Rate-Distortion Analysis}

\begin{thrm}
By using efficient lossless compression and shaping subsystems, the
distortion in encoding an \iid\ sequence generated according
to $\psrc$ can be made to approach $E[\dist{\rvSrc}{\rvQuant}]$
as closely as desired.
\end{thrm}
\begin{proof}
First we note that by assumption, we can choose $\minBlockSize$ large
enough so that the probability of the
source sequence being non-typical can be made negligible.
For a typical source sequence, we can focus on how the encoder maps
$\rvSrcT{i}$ to $\rvQuantT{i}$ since the decoder simply maps a bit
sequence to the $\rvQuantT{i}$ sequence chosen by the encoder.  The
encoder maps blocks of source samples, $\rvSrcTSub{\ct}{\ct+\cl}$, to
blocks of quantized samples, $\rvQuantTSub{\ct}{\ct+\cl}$, by using
an admissible shaping algorithm.  As described in
\secref{sec:shaping-subsystem}, the shaper produces a
$\delta$-strongly typical sequence.  
Thus the total expected
distortion is at most
\begin{equation}
E[\dist{\rvSrc}{\rvQuant}] + \dmax\cdot\delta +
\dmax\cdot\Pr[\rvSrcTSub{1}{\blockSize} \textnormal{ not typical}] 
\end{equation}
where the first two terms are the distortion for a typical sequence
produced by the shaper and the last term is the contribution from a
non-typical source sequence.
\end{proof}

\begin{thrm}
By using efficient lossless compression and shaping subsystems, the
rate in encoding an \iid\ sequence generated according
to $\psrc$ can be made to approach $I(\rvSrc;\rvQuant)$
as closely as desired.
\end{thrm}
\begin{proof}
Imagine that the parameter $\nominalBlockSize$ is chosen so that
${\numStages}$ passes of the loop in the encoding algorithm 
are executed.  Also, let $\cl_j$ denote the value of $\cl$ in
line~3 of the encoder in the $j$th pass.
We know $\cl_1 = \minBlockSize$
by construction.  By definition of an admissible shaping system in
\secref{sec:shaping-subsystem} and line~6 of
the encoder we have that $\cl_{j+1} \geq
\cl_j\cdot[H(\rvQuant)/H(\rvQuant|\rvSrc)]$.  Using this
relation and assuming that each block of length $\cl_j$ is typical,
we can compute the total number of samples encoded via
\begin{equation}
\label{eq:encoder:length}
\blockSize = \sum_{j=1}^{\numStages} \cl_j \geq 
	   \sum_{j=0}^{{\numStages}-1}  
	   \minBlockSize \cdot
	   \bhFactor^j   
= \minBlockSize\cdot\frac{[H(\rvQuant)/H(\rvQuant|\rvSrc)]^{{\numStages}}-1}{H(\rvQuant)/H(\rvQuant|\rvSrc)-1}.
\end{equation}

The bit rate required to encode these samples is 
\begin{equation}
\label{eq:encoder:bits}
R = \cl_{\numStages} \cdot H(\rvQuant) + \epsilon \leq \minBlockSize\cdot
H(\rvQuant)\cdot[H(\rvQuant)/H(\rvQuant|\rvSrc)]^{{\numStages}-1} + \epsilon\cdot {\numStages}\cdot[H(\rvQuant)/H(\rvQuant|\rvSrc)]^{\numStages}.
\end{equation}
This follows by the assumption that the admissible lossless
compression system in \secref{sec:lossl-compr-subsyst} requires
$\genericLength\cdot H(\rvQuant) + \epsilon$ bits to encode a block of
$\genericLength$ typical samples.

Therefore the number of bits per sample when the source blocks are
typical is obtained
by dividing \eqref{eq:encoder:bits} by \eqref{eq:encoder:length} to
obtain
\begin{align}
R/n &\leq \left\{\minBlockSize\cdot
H(\rvQuant) \cdot
\bhFactor^{\numStages-1}  
  + \epsilon\cdot 
\numStages
\cdot\bhFactor^{\numStages}
\right\}\bigg/\left\{\minBlockSize\cdot\frac{[H(\rvQuant)/H(\rvQuant|\rvSrc)]^{\numStages}-1}{H(\rvQuant)/H(\rvQuant|\rvSrc)-1}\right\}
\\
%
\iffalse
&=
\left\{H(\rvQuant)\bhFactor^{\numStages-1}
  \left[\hFactor-1\right] + 
\frac{\epsilon\numStages}{\minBlockSize}
  \bhFactor^{\numStages}
  \left[\hFactor-1\right]\right\}\bigg/\left\{\bhFactor^{\numStages}-1\right\}
\fi
%
&= \left\{H(\rvQuant)\left[1-\hFactorInverse\right] +
\frac{\epsilon\numStages}{\minBlockSize}\left[\hFactor-1\right]\right\}
\bigg/\left\{1-\bhFactor^{-\numStages}\right\}
\\
&=
I(\rvQuant;\rvSrc)\cdot\left\{1 + \frac{\epsilon\numStages}{\minBlockSize
  H(\rvQuant|\rvSrc)}\right\} \bigg/\left\{1-\bhFactor^{-\numStages}\right\}.
\end{align}
An extra term must also be added to account for the possibility that
the source is atypical.  By assumption we can choose
$\nominalBlockSize$ so that $\numStages$ is large enough to make
the second term in braces negligible, and then we can choose
$\minBlockSize$ so that the probability of any source block being
typical is negligible.  Also, by making $\minBlockSize$ large enough
we can make the first term in curly braces negligible.  Thus the bit
rate can be made as close to $I(\rvQuant;\rvSrc)$ as desired.

\end{proof}

Combining the previous theorems indicates that we can approach the
feedforward rate-distortion function with only the complexity required
for lossless compression and shaping systems.
\begin{corol}
When linear complexity admissible lossless compression and shaping
systems are used, the resulting feedforward rate-distortion
function can be approached arbitrarily closely with
linear complexity.
\end{corol}
In particular, we can use the lossless compression and shaping systems
described in \cite{Ooi:1998:it} which are based on arithmetic coding
and the dual of arithmetic coding respectively.

\section{Concluding Remarks}
\label{sec:concluding-remarks}

In this paper we describe a lossy compression algorithm to encode a
finite-alphabet source in the presence of feedforward information.
In particular, we show that although memoryless feedforward does not
change the rate-distortion function, it allows us to construct a low
complexity lossy compression system which approaches the
rate-distortion function.  In practice, the particular scheme
described here may require modifications and other methods of using
feedforward information or similar knowledge may be more appropriate.
Our main goal therefore is not necessarily to advocate a particular
scheme but to show that when compression, observation, and
control interact, additional resources such as feedforward may provide
advantages not available in the classic compression framework.
One interesting possibility for future work includes studying the
general problem in \secref{sec:problem-description} when the fixed lag
side information, $\rvSinfoDec$, is not exactly the same as the 
source.  Similarly, investigating the effects of memory in the source
and different values for the delay, $\FFDelay$, would also be valuable.

\appendix
\section{Information/Operational R(D) Equivalence}

\begin{prop}
The information/operational feedforward rate-distortion functions
are equal.
\end{prop}
\begin{proof}
Since the decoder must deterministically produce $\rvQuantT{i}$ from
$\rvSrcTSub{1}{i-1}$ and
the $\blockSize\rate$ bits produced by the encoder we have
\begin{align*}
\blockSize\rate &\geq \sum_{i=1}^{\blockSize}
H(\rvQuantT{i}|\rvSrcTSub{1}{i-1})
\geq \sum_{i=1}^{\blockSize} \left[
H(\rvQuantT{i}|\rvSrcTSub{1}{i-1}) -H(\rvQuantT{i}|\rvSrcTSub{1}{i})\right]
= \sum_{i=1}^{\blockSize} \left[
H(\rvSrcT{i}|\rvSrcTSub{1}{i-1}) -
H(\rvSrcT{i}|\rvSrcTSub{1}{i-1},\rvQuantT{i})\right]\\ 
%
%\label{eq:converse:memless}
&\stackrel{(a)}{=} \sum_{i=1}^{\blockSize} \left[
H(\rvSrcT{i}) -
H(\rvSrcT{i}|\rvSrcTSub{1}{i-1},\rvQuantT{i})\right]
%
%\label{eq:converse:condred}
\stackrel{(b)}{\geq}
\sum_{i=1}^{\blockSize} \left[
H(\rvSrcT{i}) -
H(\rvSrcT{i}|\rvQuantT{i})\right]
\end{align*}
where (a) follows since the source is
memoryless and (b) follows since conditioning
reduces entropy.  From this point standard convexity arguments
establish that \eqref{eq:frdfi-def} is a lower bound to the average rate.
\end{proof}

\begin{small}
\bibliographystyle{ieeetr}
\bibliography{paper}
\end{small}
\end{document}